\documentclass[aps,prl,twocolumn,superscriptaddress]{revtex4-2}

\usepackage{subfigure}
\usepackage{graphicx}
\usepackage{dcolumn}

\usepackage{bm}
\usepackage{xcolor}
\usepackage{float}
\usepackage[figuresright]{rotating}
\usepackage{longtable}
\usepackage{array}
\usepackage{csvsimple}
\usepackage{comment}
\usepackage{hyperref}
\usepackage{verbatim}
\usepackage{lineno}
\usepackage{multirow}
\usepackage{enumitem} 
\usepackage{amsmath}
\definecolor{aa}{RGB}{0,0,139}

\newcommand{\chisq}{\chi^2}
\begin{document}

\preprint{APS/123-QED}

\title{{\bf \boldmath Precise measurement of the $e^+e^-\to D_s^+D_s^-$ cross sections at center-of-mass energies from threshold to 4.95~GeV}}

\author{M.~Ablikim$^{1}$, M.~N.~Achasov$^{4,c}$, P.~Adlarson$^{75}$, O.~Afedulidis$^{3}$, X.~C.~Ai$^{80}$, R.~Aliberti$^{35}$, A.~Amoroso$^{74A,74C}$, Q.~An$^{71,58,a}$, Y.~Bai$^{57}$, O.~Bakina$^{36}$
% ...
\\
\vspace{0.2cm}
(BESIII Collaboration)\\
\vspace{0.2cm} {\it
$^{1}$ Institute of High Energy Physics, Beijing 100049, People's Republic of China\\
% ...
$^{3}$ Bochum  Ruhr-University, D-44780 Bochum, Germany\\
$^{4}$ Budker Institute of Nuclear Physics SB RAS (BINP), Novosibirsk 630090, Russia\\
% ...
$^{35}$ Johannes Gutenberg University of Mainz, Johann-Joachim-Becher-Weg 45, D-55099 Mainz, Germany\\
$^{36}$ Joint Institute for Nuclear Research, 141980 Dubna, Moscow region, Russia\\
% ...
$^{57}$ Southeast University, Nanjing 211100, People's Republic of China\\
$^{58}$ State Key Laboratory of Particle Detection and Electronics, Beijing 100049, Hefei 230026, People's Republic of China\\
% ...
$^{71}$ University of Science and Technology of China, Hefei 230026, People's Republic of China\\
% ...
$^{74}$ University of Turin and INFN, (A)University of Turin, I-10125, Turin, Italy; (B)University of Eastern Piedmont, I-15121, Alessandria, Italy; (C)INFN, I-10125, Turin, Italy\\
$^{75}$ Uppsala University, Box 516, SE-75120 Uppsala, Sweden\\
% ...
$^{80}$ Zhengzhou University, Zhengzhou 450001, People's Republic of China\\
\vspace{0.2cm}
$^{a}$ Deceased\\
% ...
$^{c}$ Also at the Novosibirsk State University, Novosibirsk, 630090, Russia\\
% ...
}}


\vspace{0.4cm}
\date{\today}

\begin{abstract}
Using the $e^+e^-$ collision data collected with the BESIII detector operating at the BEPCII collider, at center-of-mass energies from the threshold to $4.95$~GeV, we present precise measurements of the cross sections for the process $e^+e^-\to D_s^+D_s^-$ using a single tag method. The resulting cross section lineshape exhibits several new structures, thereby offering an input for coupled channel analysis and model tests, which are critical to understand vector charmonium-like states with masses between 4 and 5~GeV.
\end{abstract}

\pacs{Valid PACS appear here}
\maketitle

Quantum Chromodynamics (QCD) predicts the confinement phenomenon in the low-energy region where hadrons are formed. A detailed study of hadron systems is essential for a better understanding of the non-perturbative effect of QCD. An important milestone in hadron physics is the discovery of the vector charmonium-like states, such as the $\psi(4230)$ in the $e^+e^-\to \pi^+\pi^-J/\psi$ process~\cite{ISRY4260ToPiPiJpsi-BaBar,ISRY4260ToPiPiJpsi2007-Belle}, and the $\psi(4360)$ and $\psi(4660)$ in the $e^+e^-\to \pi^+\pi^-\psi(3686)$ process~\cite{ISRPipiPsipBelle,ISRPipiPsipBabar}; These states exhibit properties that do not match the expectations for pure $c\bar{c}$ states. Since the masses of these states lie above the open charm threshold, measurement of their couplings to the open-charm channels is crucial for understanding of their nature. However, such studies are very complicated due to the presence of coupled channel effects. The knowledge of exclusive open-charm cross sections, such as the cross sections of $e^+e^-\to D_s^+D_s^-$~\cite{Uglov:2016orr}, is highly desirable. 

Although the Belle and BaBar experiments have measured the exclusive cross sections for $e^+e^-\to D_s^+D_s^-$ through initial state radiation (ISR) processes~\cite{ISR_Belle,ISR_BaBah}, the precision is not sufficient to pin down the correlation between the $\psi(4230)$ and the process $e^+e^-\to D_s^+D_s^-$. The CLEO-c experiment studied the process $e^+e^-\to D_s^+D_s^-$~\cite{cleoc} using the energy scan data, with the maximum center-of-mass energy ($E_{\rm cm}$) of only 4.26 GeV.

The BESIII detector~\cite{BESIIIdetector,whitepaper} has collected larger data samples over a broader energy range, which allow improved measurements of the exclusive open-charm cross sections. One example is the recent publication of the $e^+e^-\to D_s^{*+}D_s^{*-}$ cross sections, where an unusual cross section lineshape is observed~\cite{BESIII:2023wsc}. 

In this Letter, the cross sections of $e^+e^-\to D_s^+D_s^-$ are measured in the $E_{\rm cm}$ range 
between 3.94 GeV and $4.95$~GeV with 138 energy points corresponding to an integrated luminosity 
of $22.9~\textrm{fb}^{-1}$~in total. The so-called $XYZ$ data samples (collected for the study of $XYZ$ states) account for 95\% of the total integrated luminosity. The remaining data samples, referred to as the $R$-scan data samples (collected for the $R$ value measurement), have an integrated luminosity around $7~\textrm{pb}^{-1}$~at each energy point. 

To optimize the event selection criteria, determine the detection efficiency, and estimate the backgrounds, we produce simulated samples using the  {\scriptsize GEANT4}-based~\cite{geant4} Monte Carlo (MC) software, which includes the geometric description of the BESIII detector and its response. The signal MC samples of the $e^+e^-\to D_s^+D_s^-$ process are produced for each energy point, where the vector state decays into two scalar particles are generated by the EvtGen~\cite{vss}, and the $D_{s}^{+}\to K^+K^-\pi^+$ are generated with the Dalitz model~\cite{dalitz1,dalitz2}. The beam energy spread and ISR are taken into account with the generator~{\scriptsize \textsc{KKMC}}~\cite{kkmc1,kkmc2}. To estimate possible background contributions, we generate inclusive and exclusive MC samples modeled with EvtGen using branching fractions taken from the Particle Data Group (PDG)~\cite{PDG} for the known decay modes, and with {\sc lundcharm}~\cite{ref:lundcharm} for the remaining unknown charmonium decays. The integrated luminosities of the inclusive samples are comparable to those of data. Final-state radiation from charged particles is incorporated with the {\sc photos} package~\cite{photos}.  

Due to the low background level and high detection efficiency at BESIII, 
we only reconstruct the $D_s^- \rightarrow K^+K^-\pi^-$ decay, 
while $D_s^+$ is tagged by the recoil mass. The charge conjugated modes are implied throughout this Letter, 
and the tagged $D_s$ will always be referred to as $D_s^-$. 
Selection and identification of charged particles are performed
using the same criteria as described in Ref.~\cite{eventselection}. 
An event is selected as a single tag candidate if it contains at least one $K^+K^-$ pair
and at least one charged pion. 
All single tag candidates are kept for further analysis. To improve the signal purity, we select the $D_s^-$ candidates with two intermediate decay modes:  
\mbox{$D_s^-\to\phi\pi^-$}~with~\mbox{$\phi\to K^+K^-$} or \mbox{$D_s^-\to K^{*}(892)^{0}K^-$}~with~\mbox{$K^{*}(892)^0\to K^+\pi^-$}. The $D_s^-\to\phi\pi^-$ decays are selected requiring the invariant mass of $K^+K^-$ to satisfy \mbox{$1.005<M(K^+K^-)<1.035~{\rm GeV}/c^2$}. 
To select the $D^-_s~\to ~K^{*}(892)^0K^-$ decays,
we apply the requirements on the $K^+\pi^-$  invariant mass of \mbox{$0.832<M(K^+\pi^-)<0.928$~GeV/$c^2$} and 
on the helicity angle of \mbox{$\lvert \cos\theta_{K^+\text{ in }K^+\pi^-}\rvert < 0.52$}. 
The helicity angle is defined as the angle between the $K^*(892)^0$ momentum in the $D_s^-$ 
rest frame and the $K^+$ momentum in the $K^*(892)^0$ rest frame.

\begin{figure*}[htbp]
 \centering
 \includegraphics[width=0.9\textwidth]{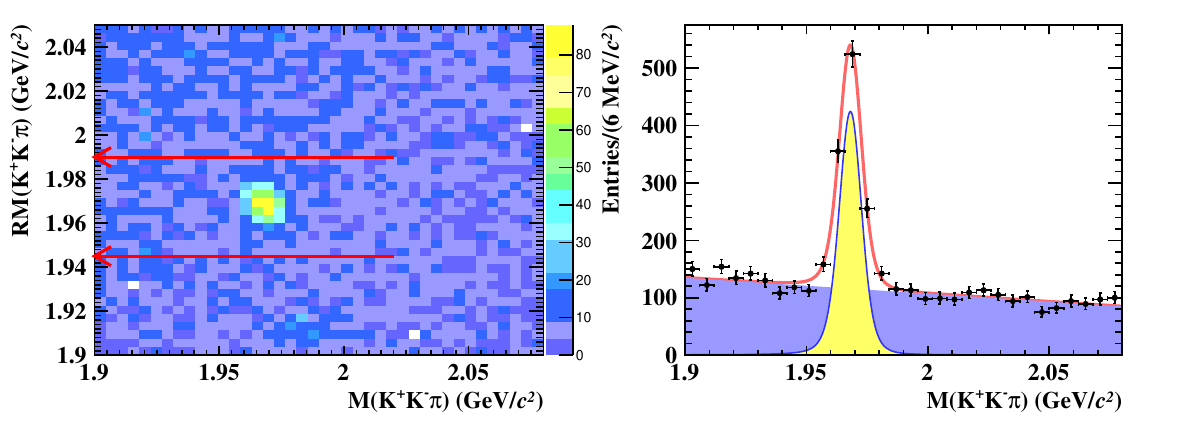}
  \includegraphics[width=0.9\textwidth]{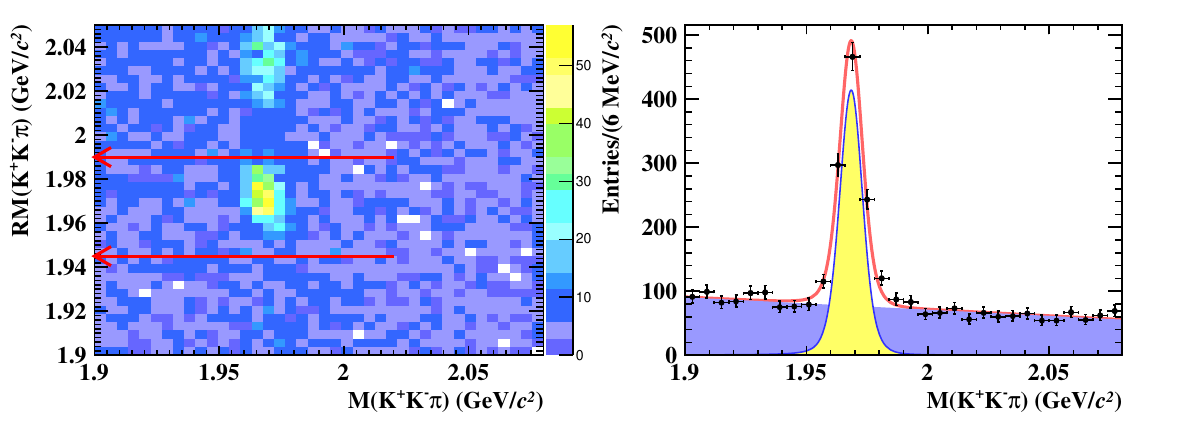}
\caption{The distribution plot of $RM(D_s^{\pm})$ versus $M(K^+K^-\pi^{-})$ (left) and the fit to $M(K^+K^-\pi^{-})$ (right) for data at $E_{\rm cm}=4.26$(first row)~and 4.68(second row)~GeV. The signal interval is indicated by the red arrows. The fit to the $M(K^+K^-\pi)$ distribution is shown in the right panels as a red curve, with the background shape in blue and signal shape in yellow. 
}\label{pic:figure1}
 \end{figure*}

Figure~\ref{pic:figure1} (left) shows the scatter plot of the invariant mass of the selected $K^+K^-\pi$ combination ($M(K^+K^-\pi)$) versus its recoil mass ($RM(D_s^-)$) for data at \mbox{$E_{\rm cm}=4.26$~and 4.68~GeV}. A clear cluster of events corresponding to the $D_s^+D_s^-$ pair is observed. Similar signals are observed in the data samples at other energy points. The resolution of the recoil mass 
is improved by defining \mbox{$RM(D_s^-)\equiv M^{\rm recoil}_{K^+K^-\pi^-}+~M(K^+K^-\pi^-)-m(D_s^-)$}, 
where \mbox{$M^{\rm recoil}_{K^+K^-\pi^-}\equiv \vert \vec p_{e^+e^-}-\vec p_{K^+}-\vec p_{K^-}-\vec p_{\pi^-}\vert$}, 
\mbox{$\vec p_{e^+e^-}$, $\vec p_{K^+}$, $\vec p_{K^-}$, and $\vec p_{\pi^-}$} are the four-momenta of the initial $e^+e^-$ system, the selected $K^+$, $K^-$, and $\pi^-$, respectively, and $m(D_s^-)$ is the nominal mass of $D_{s}^{-}$~\cite{PDG}. To suppress the background, the $RM(D_s^-)$ is required to be within the expected signal window, 
$1.945< RM(D_s^-)<1.990~{\rm GeV}/c^2$, as shown in Fig.~\ref{pic:figure1}(left).

In order to obtain the signal yields, we fit the $M(K^+K^-\pi)$ distribution after applying all the above requirements. The signal shape is described with the signal MC samples convolved with a Gaussian function. 
The differences between MC simulation and data are taken into account 
with the mass shift $\Delta m$ and the resolution $\Delta \sigma$ of the Gaussian function. 
At all energy points, $\Delta \sigma$ is fixed to the average value of 2.4 MeV to improve the fit stability 
for the low-statistics samples, while $\Delta m$ is a free parameter of the fit at each energy point.
The background is described with a linear function. The fits to $M(K^+K^-\pi)$ for data at $E_{\rm cm}=4.26$~GeV and 4.68 GeV are shown in Fig.~\ref{pic:figure1}(right) as examples. 

Studies of the inclusive MC samples show that the $e^+e^-\to D_s^+D_s^{*-}$ process contributes as a source of peaking background at $E_{\rm cm}>4.6$~GeV. The estimated numbers of the $D_s^+$ signal events from $e^+e^-\to D_s^+D_s^{*-}$ (defined as $N_{D_s^{\pm}D_s^{\mp\ast}}$) are obtained
from exclusive MC samples normalized to the corresponding integrated luminosity and cross sections, and are subtracted when calculating the number of signal events. The process $e^+e^- \to D_{s}^{*+}D_{s}^{*-}$ fails the recoil mass requirement, as its $RM(D_s^-)$ is concentrated above 2.1~GeV, making the signal contamination by this background negligible.

The Born cross section of $e^+e^-\to D_s^+D_s^-$ is determined with
\begin{equation}\label{equ:xsection}
\begin{aligned}
    \sigma_{\rm Born}&= \frac{ N_{D_s}^{\rm fit} - N_{D_s^{\pm}D_s^{\mp\ast}}}{2\mathcal{B}(D_s^\pm\to K^+K^- \pi^\pm) \epsilon(1+\delta)\frac{1}{|1-\Pi|^{2}}\mathcal{L}},
\end{aligned}
\end{equation}
where $N_{D_s}^{\rm fit}$ is obtained from the fit to the $M(K^+K^-\pi^{\pm})$ distribution in data.
The factor 1/2 takes into account the contributions from both $D_s^+$ and $D_s^-$ single tag reconstruction. The $N_{D_s^{\pm}D_s^{\mp*}}$ is the peaking background from $e^+e^-\to D_s^+D_s^{*-}$, which is only present at $E_{\rm cm}>4.6$~GeV; $\mathcal{L}$ is the integrated luminosity, $(1+\delta)$ is the ISR correction factor, $\frac{1}{|1-\Pi|^{2}}$ is the vacuum polarization (VP)~\cite{vpcorraction} factor, and $\mathcal{B}(D_s^\pm\to K^+K^-\pi^\pm)$ is the branching fraction of $D_s^\pm\to K^+K^-\pi^\pm$, which is taken from Ref.~\cite{PDG}; $\epsilon$ is the detection efficiency.

To obtain the Born cross sections, we use an MC-weighting method for radiative corrections and the efficiencies, in which the correction factors are evaluated iteratively. Initially, $(1 + \delta)$ and $\epsilon$ are first obtained by using {\scriptsize \textsc{KKMC}} simulation with a flat Born cross section lineshape as input. 
Once the initial cross section values are obtained, a smooth lineshape can be derived using a variable span smoother~\cite{super1,super2}. By utilizing the smoothed lineshape, the MC-weighting method~\cite{Sun:2020ehv} is employed to update the efficiencies, ISR correction factors, and measured cross sections. After five iterations, the Born cross sections converge to the values shown in Fig.~\ref{pic:cs_Born}. The detailed numbers are given in the Supplemental Material~\cite{material1}. 

\begin{figure*}[htbp]
 \centering
 \includegraphics[width=1.05\textwidth]{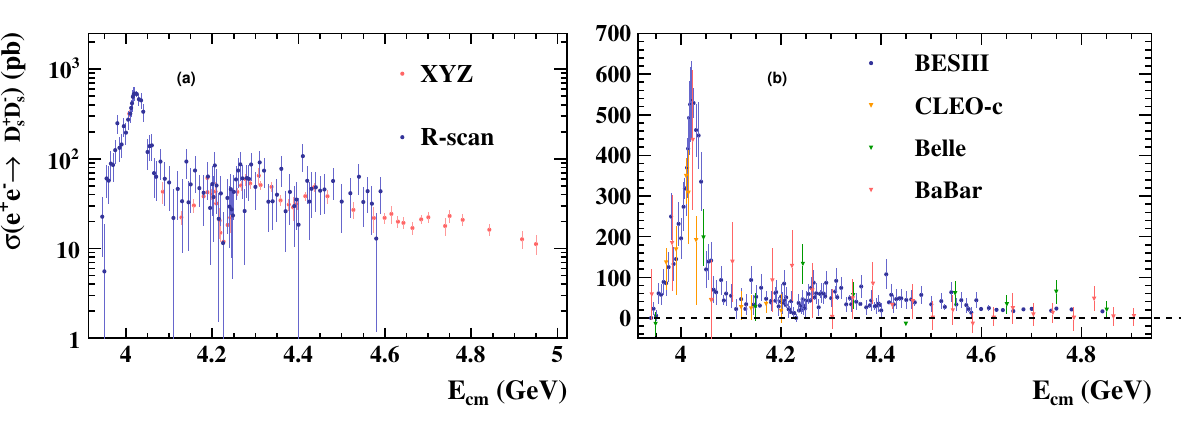}
 \caption{The measured Born cross sections of $e^+e^-\to D_s^+D_s^-$ in logarithmic coordinates (a), where the error bars include the statistical and systematic uncertainties. The comparison between our measurements and those of BaBar, Belle, and CLEO-c (after ISR correction) is shown in (b). }\label{pic:cs_Born}
 \end{figure*}

Several features are observed in Fig.~\ref{pic:cs_Born}(a).
First, the cross section peak above the $D_s^+D_s^-$ threshold appears around $E_{\rm cm}=4.02$ ~GeV, which is close to the mass
of the known $\psi(4040)$. However, the width of this structure, estimated with a Breit-Wigner function fit, 
is about 40$\pm \rm 4$~MeV, narrower than the $80\pm10$~MeV width of $\psi(4040)$ listed in the PDG~\cite{PDG}. 
This narrower width implies the presence of a strong coupled channel effect as predicted in Ref.~\cite{Eichten:1979ms}. 
The strong coupled channel effect may help to explain many phenomena observed above the open charm threshold. 
The measurements reported here are important for the validation of this theory.

The measured maximum cross sections around 4.02~GeV are approximately 500~pb, exceeding the previous measurements, including those from Belle, BaBar, and CLEO-c after ISR correction~\cite{Wangll}, as shown in Fig.~\ref{pic:cs_Born}(b). One possible explanation for the discrepancy is that the bin sizes used in the previous measurements are too large to discern this narrow peak.

The second notable feature is a narrow dip observed around $E_{\rm cm}=4.23$~GeV, which is close to the $D_s^{*+}D_s^{*-}$ threshold ($4.224$~GeV) and the peak of the $\psi(4230)$. This phenomenon may reflect the influence of the open channel effect of $e^+e^-\to D_s^{*+}D_s^{*-}$ on the cross-section lineshape of $e^+e^-\to D_s^+D_s^-$. It is important to note that the decay width of $\psi(4230)\to f_0(980) J/\psi$~\cite{pipijpisi_bes, pipijpsi_belle3} is not negligible, and the $f_0(980)$ particle contains a significant $s\bar s$  component~\cite{f980}. Since the $s\bar s$ quark pair is also present in the $e^+e^-\to D_s^+D_s^-$ process, the observation of this dip offers valuable insights into the aforementioned nature of the $\psi(4230)$. If we disregard this dip, another notable observation is the presence of a broad structure spanning from $E_{\rm cm}=$ 4.1 to 4.4~GeV. 

\begin{figure}[htbp]
 \centering
 \includegraphics[width=0.5\textwidth]{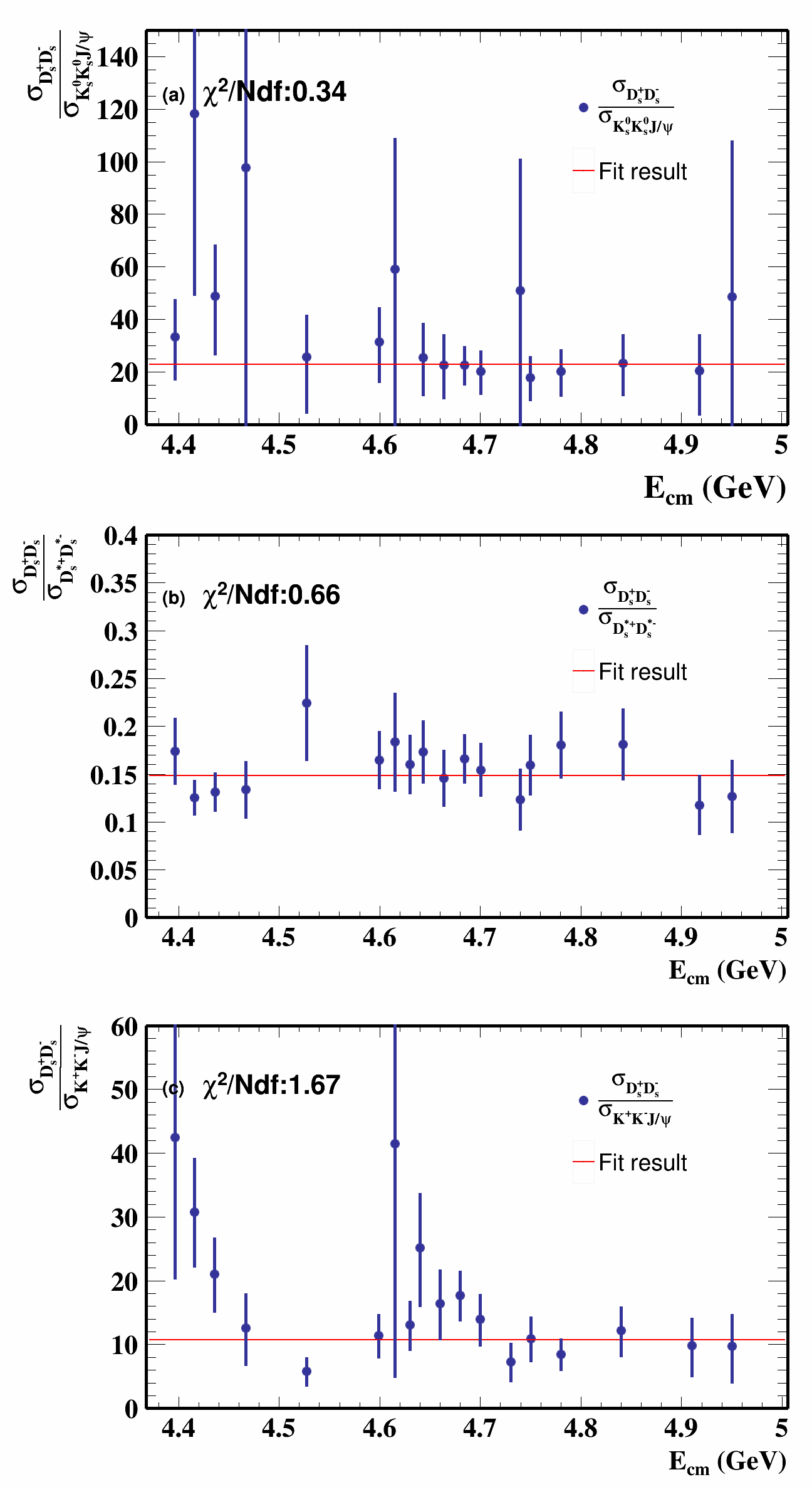}
 \caption{The Born cross section ratios between $e^+e^-\to D_s^+D_s^-$ and $e^+e^-\to K^{0}_{S}K^{0}_{S}J/\psi$(a). The Born cross section ratios between $e^+e^-\to D_s^+D_s^-$ and $e^+e^-\to D_s^{*+}D_s^{*-}$(b). The Born cross section ratios between  $e^+e^-\to D_s^+D_s^-$ and $e^+e^-\to K^+K^-J/\psi$(c). Only statistical uncertainties are included.}\label{pic:factor}
 \end{figure}

Particularly, our measured lineshape exhibits similarities in structures with the previously reported $e^+e^-\to D_s^{*+}D_s^{*-}$~\cite{BESIII:2023wsc}, $e^+e^-\to K^+K^-J/\psi$~\cite{kkjpsi1,kkjpsi2}, and $e^+e^-\to K_S^{0}K_S^{0}J/\psi$~\cite{ks0ks0jpsi} processes above $E_{\rm cm}=4.4$~GeV. The Born cross section ratios $\frac{\sigma(e^+e^+\to D_s^+ D_s^-)}{e^+e^-\to K_S^{0}K_S^{0}J/\psi}$, $\frac{\sigma(e^+e^+\to D_s^+ D_s^-)}{\sigma(e^+e^-\to D_s^{*+}D_s^{*-})}$, and $\frac{\sigma(e^+e^+\to D_s^+ D_s^-)}{\sigma(e^+e^-\to K^+K^-J/\psi})$ are shown in Fig.~\ref{pic:factor}. Fits to the ratios with constant values give $\chisq/{\rm ndf}$=0.37, 1.09, and 1.69, respectively, where ${\rm ndf}$ is the number of degrees of freedom. In the case of $e^+e^-\to K^+K^- J/\psi$, two new structures have been observed around $E_{\rm cm}=$~4.5 and 4.7~GeV~\cite{kkjpsi2}, while the fit result for $e^+e^-\to D_s^{*+}D_s^{*-}$ suggests a new possible structure around $E_{\rm cm=}$~4.79~GeV with a significance of $6\sigma$~\cite{BESIII:2023wsc}. The good agreement in structure and similarity in quark composition with final state particles make $e^+e^-\to D_s^+D_s^-$ a promising candidate process for further investigation of the observed structures.

The systematic uncertainty of the cross section measurement includes two groups of contributions: independent and common for all energy points.

The common systematic uncertainties include uncertainties related to the integrated luminosity, tracking, particle identification (PID), fit range, and cross section lineshape description as shown in Table~\ref{tab:sysSum1}. The integrated luminosities are measured with Bhabha and digamma events~\cite{XYZlumi} with a systematic uncertainty of 1.0\%. The PID is performed on two kaons and one pion and the corresponding systematic uncertainty is 1.0\% per track~\cite{PID}. The uncertainty of tracking efficiency is estimated to be 1.0\% per track~\cite{track}. Therefore, a 3.0\% systematic uncertainty is assigned for both the tracking efficiency and PID. The uncertainty due to the fit range is estimated to be 2.0\% by varying the mass range by 10~MeV. Since ISR correction depends on the dressed cross section line shape, the difference of smoothed lineshapes estimated with and without statistical uncertainties is considered as its systematic uncertainty of 2\%.
\begin{table}[htpb]
 \centering
  \caption{Common relative systematic uncertainties in the measured Born cross sections. }\label{tab:sysSum1} 
  \resizebox{0.25\textwidth}{!}{
 \begin{tabular}{c|c}
    \hline \hline
    Source & $\sigma_\textrm{sys}$ (\%) \\
    \hline
    Tracking & 3.0\\
    PID &3.0 \\
    $\mathcal{B}(D_s^\pm\to K^+K^-\pi^\pm)$ & 1.9\\
    Integrated luminosity & 1.0 \\
    Fit range &2.0\\
    Lineshape &2.0\\
    \hline 
    Total & 5.5 \\
    \hline \hline
\end{tabular}
}
\end{table}

We only evaluate energy-dependent systematic uncertainties at 28 high-statistics energy points from the $XYZ$ samples. 
The systematic uncertainties for the samples with low statistics are estimated using the values obtained from those nearby energy points with high statistics. We consider four sources of energy-dependent systematic uncertainties, including two mass windows and the choice of the signal and background shapes used in the fits. 
The uncertainties of the mass windows of $\phi$ and $K^{*0}$ arise due to the difference in mass resolution between data and MC simulation. 
To evaluate the effect, the invariant mass distribution in the MC is smeared by a Gaussian distribution
to obtain an agreement in mass resolution with the data. The relative change of the selection efficiency
due to the smearing is taken as the systematic uncertainty.
The uncertainty due to the signal shape is estimated by comparing fit results when the $\Delta(\sigma)$ parameter
is fixed to 2.4 MeV and when it is used as a free parameter of the fit. 
The uncertainty due to the background shape is estimated by 
describing the background with a second-order polynomial instead of a linear function. The uncertainty due to the vacuum polarization is negligible.  The energy-dependent systematic uncertainties range from 1.1\% to 4.2\%, dominated by the contribution from the signal shape.

The total systematic uncertainties are calculated by summing all individual items in quadrature, which vary within (5.6-6.9)\%. All the systematic uncertainties, including the individual contributions and the total ones, are presented as a covariance matrix in the Supplemental Material~\cite{material3}.

In summary, we present precise measurements of the cross sections for the process $e^+e^-\to D_s^+D_s^-$ for $E_{\rm cm}$ ranging from 3.94 to 4.95~GeV. Several notable features are observed in the cross section lineshape. Firstly, a significantly narrower width of the $\psi(4040)$ than the PDG value is observed. Secondly, a dip around the $D_s^{*+}D_s^{*-}$ threshold and the peak position of the $\psi(4230)$ is observed, suggesting the influence of the open channel effect. Additionally, two structures, which are similar to those observed in $e^+e^-\to D_s^{*+}D_s^{*-}$, $e^+e^-\to K^+K^-J/\psi$, and $e^+e^-\to K_S^{0}K_S^{0}J/\psi$, are identified. These are critical for deeper understanding of the conventional and exotic states in this energy region. 

Instead of performing a fit to the cross section lineshape using conventional Breit-Wigner functions, we provide the cross section values, their uncertainties, and a correlation matrix directly for further investigation. This is motivated by the challenges in accurately fitting the lineshape when strong coupled-channel effects are present. These measurements open up an new avenue to test models explaining the nature of the $\psi(4230)$ and other vector charmonia and charmonium-like states as well as perform coupled-channel fits involving multiple channels. 

The BESIII Collaboration thanks the staff of BEPCII and the IHEP computing center for their strong support. This work is supported in part by National Key R\&D Program of China under Contracts Nos. 2020YFA0406300, 2020YFA0406400, 2023YFA1606000; National Natural Science Foundation of China (NSFC) under Contracts Nos. 11635010, 11735014, 11835012, 11935015, 11935016, 11935018, 11961141012, 12025502, 12035009, 12035013, 12061131003, 12192260, 12192261, 12192262, 12192263, 12192264, 12192265, 12221005, 12225509, 12235017, 12150004; Program of Science and Technology Development Plan of Jilin Province of China under Contract No. 20210508047RQ and 20230101021JC; the Chinese Academy of Sciences (CAS) Large-Scale Scientific Facility Program; the CAS Center for Excellence in Particle Physics (CCEPP); Joint Large-Scale Scientific Facility Funds of the NSFC and CAS under Contract No. U1832207; CAS Key Research Program of Frontier Sciences under Contracts Nos. QYZDJ-SSW-SLH003, QYZDJ-SSW-SLH040; 100 Talents Program of CAS; The Institute of Nuclear and Particle Physics (INPAC) and Shanghai Key Laboratory for Particle Physics and Cosmology; European Union's Horizon 2020 research and innovation programme under Marie Sklodowska-Curie grant agreement under Contract No. 894790; German Research Foundation DFG under Contracts Nos. 455635585, Collaborative Research Center CRC 1044, FOR5327, GRK 2149; Istituto Nazionale di Fisica Nucleare, Italy; Ministry of Development of Turkey under Contract No. DPT2006K-120470; National Research Foundation of Korea under Contract No. NRF-2022R1A2C1092335; National Science and Technology fund of Mongolia; National Science Research and Innovation Fund (NSRF) via the Program Management Unit for Human Resources \& Institutional Development, Research and Innovation of Thailand under Contract No. B16F640076; Polish National Science Centre under Contract No. 2019/35/O/ST2/02907; The Swedish Research Council; U. S. Department of Energy under Contract No. DE-FG02-05ER41374


\begin{thebibliography}{40}

\bibitem{ISRY4260ToPiPiJpsi-BaBar} \href{https://journals.aps.org/prl/abstract/10.1103/PhysRevLett.95.142001}{B. Aubert \emph{et al.} (BaBar Collaboration), Phys. Rev. Lett. \textbf{95}, 142001 (2005).}

\bibitem{ISRY4260ToPiPiJpsi2007-Belle} \href{https://journals.aps.org/prl/abstract/10.1103/PhysRevLett.99.182004}{C. Z. Yuan \emph{et al.} (Belle Collaboration), Phys. Rev. Lett. \textbf{99}, 182004 (2007).}

\bibitem{ISRPipiPsipBelle} \href{https://journals.aps.org/prl/abstract/10.1103/PhysRevLett.99.142002}{X. L. Wang \emph{et al.} (Belle Collaboration), Phys. Rev. Lett. \textbf{99}, 142002 (2007).}

\bibitem{ISRPipiPsipBabar} \href{https://journals.aps.org/prd/abstract/10.1103/PhysRevD.89.111103}{J. P. Lees \emph{et al.} (BaBar Collaboration), Phys. Rev. D \textbf{89}, 111103(R) (2014).}

\bibitem{Uglov:2016orr}\href{https://arxiv.org/abs/1611.07582}{
T.~V.~Uglov, Y.~S.~Kalashnikova, A.~V.~Nefediev, G.~V.~Pakhlova and P.~N.~Pakhlov,
JETP 1-7 (2017).}

\bibitem{ISR_Belle} \href{https://journals.aps.org/prd/abstract/10.1103/PhysRevD.83.011101}{G. Pakhlova \emph{et al.} (Belle Collaboration), Phys. Rev. D \textbf{83}, 011101 (2011).}

\bibitem{ISR_BaBah} \href{https://journals.aps.org/prd/abstract/10.1103/PhysRevD.82.052004}{P. del Amo Sanches \emph{et al.} (BaBar Collaboration), Phys. Rev. D \textbf{82}, 052004 (2010).}

\bibitem{cleoc} \href{https://journals.aps.org/prd/abstract/10.1103/PhysRevD.80.072001}{D. Cronin-Hennessy \emph{et al.} (CLEO Collaboration), Phys. Rev. D \textbf{80}, 072001 (2009).}

\bibitem{whitepaper} \href{https://iopscience.iop.org/article/10.1088/1674-1137/44/4/040001}{M.~Ablikim \textit{et al.} (BESIII Collaboration), Chin. Phys. C {\bf 44}, 040001 (2020).}

\bibitem{BESIIIdetector} \href{https://arxiv.org/abs/0911.4960}{M.~Ablikim {\it et al.} (BESIII Collaboration), Nucl.\ Instrum.\ Meth.\ A {\bf 614}, 345 (2010).}

\bibitem{BESIII:2023wsc} \href{https://journals.aps.org/prl/abstract/10.1103/PhysRevLett.131.151903}{M.~Ablikim {\it et al.} (BESIII Collaboration), Phys.\ Rev.\ Lett. {\bf 131}, 151903 (2023).}

\bibitem{geant4} \href{https://doi.org/10.1016/S0168-9002(03)01368-8}{S. Agostinelli \emph{et al.} (GEANT4 Collaboration), Nucl.{\bf309} Instrum. Meth. A {\bf 506}, 250 (2003).}

\bibitem{vss} \href{https://iopscience.iop.org/article/10.1088/1674-1137/32/8/001}{R. G. Ping, Chin. Phys. C \textbf{32}, 599 (2008).}
\bibitem{dalitz1}\href{https://link.aps.org/doi/10.1103/PhysRevD.88.032009}{P. U. E. Onyisi \textit{et al.} (CLEO Collaboration), Phys. Rev. D {\bf 88}, 032009 (2013).}
\bibitem{dalitz2} \href{https://link.aps.org/doi/10.1103/PhysRevD.104.012016}{M. Ablikim \textit{et al.} (BESIII Collaboration), Phys. Rev. D {\bf 104}, 012016 (2021).}
\bibitem{kkmc1} \href{https://www.sciencedirect.com/science/article/pii/S0010465500000485}{S.~Jadach, B.~F.~L.~Ward and Z.~Was, Comput. Phys. Commun., {\bf 130}, 260325 (2000).}

\bibitem{kkmc2} \href{https://link.aps.org/doi/10.1103/PhysRevD.63.113009}{S.~Jadach, B.~F.~L.~Ward and Z.~Was, Phys.\ Rev.\ D {\bf 63}, 113009 (2001).}

\bibitem{PDG} \href{https://doi.org/10.1093/ptep/ptac097}{R.~L.~Workman \textit{et al.} (Particle Data Group),Prog. Theor. Exp. Phys. \textbf{2022}, 083C01 (2022).}

\bibitem{ref:lundcharm} \href{https://link.aps.org/doi/10.1103/PhysRevD.62.034003}{J.~C.~Chen, G.~S.~Huang, X.~R.~Qi, D.~H.~Zhang and Y.~S.~Zhu, Phys.\ Rev.\ D {\bf 62}, 034003 (2000);}\href{https://iopscience.iop.org/article/10.1088/0256-307X/31/6/061301}{R.~L.~Yang, R.~G.~Ping and H.~Chen, Chin.\ Phys.\ Lett.\  {\bf 31}, 061301 (2014).}

\bibitem{photos}\href{https://doi.org/10.1016/0370-2693(93)90062-M}{E.~Richter-Was,Phys.\ Lett.\ B {\bf 303}, 163 (1993).}

\bibitem{eventselection} \href{https://link.aps.org/doi/10.1103/PhysRevD.101.112008}{M. Ablikim \emph{et al.} (BESIII Collaboration), Phys. Rev. D. {\bf 101},~112008 (2020).}

\bibitem{vpcorraction} \href{https://doi.org/10.1140/epjc/s10052-010-1251-4}{S. Actis \emph{et al.} Eur. Phys. J. C {\bf 66}, 585 (2010).}

\bibitem{super1} \href{https://api.semanticscholar.org/CorpusID:60507696}{Jerome~H.~Friedman, SMART User's Guide,Laboratory for Computational Statistics, Stanford University Technical Report No. 1. (1984).}
\bibitem{super2} \href{https://api.semanticscholar.org/CorpusID:222281453}{Jerome~H.~Friedman, A variable span scatterplot smoother. Laboratory for Computational Statistics, Stanford University Technical Report No.5.(1984).}
 
\bibitem{Sun:2020ehv} \href{https://doi.org/10.1007/s11467-021-1085-6}{W.~Sun, T.~Liu, M.~Jing, L.~Wang, B.~Zhong and W.~Song,Front. Phys. (Beijing) \textbf{16}, 64501 (2021).}

\bibitem{material1} Supplemental Material [link to be added] for the detailed numbers of the cross sections.


\bibitem{Eichten:1979ms} \href{https://link.aps.org/doi/10.1103/PhysRevD.21.203}{E.~Eichten, K.~Gottfried, T.~Kinoshita, K.~D.~Lane and T.~M.~Yan, Phys. Rev. D \textbf{21}, 203 (1980).}

\bibitem{Wangll} \href{https://dx.doi.org/10.1088/1674-1137/42/4/043002}{Xiang-Kun Dong, Liang-Liang Wang, and Chang-Zheng Yuan, Chin. Phys.C {\bf 42}, 043002 (2018).}
\bibitem{pipijpisi_bes}\href{https://journals.aps.org/prl/abstract/10.1103/PhysRevLett.111.019901}{M. Ablikim \emph{et al.} (BESIII Collaboration), Phys. Rev. Lett \textbf{110},252001(2013)}


\bibitem{pipijpsi_belle3}\href{https://journals.aps.org/prl/abstract/10.1103/PhysRevLett.110.252002}{Z. Q. Liu \emph{et al.} (Belle Collaboration), Phys. Rev. Lett. \textbf{110}, 252002 (2013).}

\bibitem{f980}\href{https://inspirehep.net/literature/534100}{ E.~van Beveren, G.~Rupp and M.~D.~Scadron, Phys. Lett. B \textbf{495}, 300-302 (2000)}
\bibitem{kkjpsi1}\href{https://link.aps.org/doi/10.1103/PhysRevD.97.071101}{M. Ablikim \emph{et al.} (BESIII Collaboration), Phys. Rev. D \textbf{97}, 071101 (2018).}

\bibitem{kkjpsi2}\href{https://link.aps.org/doi/10.1103/PhysRevLett.131.211902}{M. Ablikim \emph{et al.} ((BESIII Collaboration), Phys. Rev. Lett \textbf{131}, 211902 (2023).}

\bibitem{ks0ks0jpsi}\href{https://link.aps.org/doi/10.1103/PhysRevD.107.092005}{
M. Ablikim \emph{et al.} (BESIII Collaboration), Phys. Rev. D. {\bf 107}, {092005} (2023).}
\bibitem{XYZlumi} \href{https://dx.doi.org/10.1088/1674-1137/39/9/093001}{M. Ablikim \emph{et al.} (BESIII  Collaboration), Chin. Phys. C \textbf{39}, 093001 (2015).}
\bibitem{PID} \href{https://link.aps.org/doi/10.1103/PhysRevD.91.032002}{M. Ablikim {et al}. (BESIII Collaboration), Phys. Rev. D \textbf{91}, 032002 (2015).}
\bibitem{track} \href{https://link.aps.org/doi/10.1103/PhysRevLett.112.022001}{M. Ablikim \emph{et al.} (BESIII Collaboration), Phys. Rev. Lett. \textbf{112}, 022001 (2014).}
\bibitem{material3} Supplemental Material [link to be added] for covariance systematic uncertainties matrix.

\end{thebibliography}

\end{document}